\documentclass[fleqn,usenatbib]{mnras}

\usepackage{newtxtext,newtxmath}

\usepackage[T1]{fontenc}
\usepackage[utf8]{inputenc}
\usepackage{ae,aecompl}

\usepackage{xifthen}
\usepackage{xspace}

\usepackage{graphicx}	
\usepackage{amsmath}	

\newcommand{\Galform}{{\sc galform}}

\newcommand{\unit}[1]{\ensuremath{\mathrm{\,#1}}\xspace}
\newcommand{\unitlogicspace}[2]{%
  \ifthenelse{\isempty{#1}}%
    {\unit{#2}}
    {\ensuremath{{#1}\, \unit{#2}}}
  }
\newcommand{\kms}[1][]{\unitlogicspace{#1}{km\, s^{-1}}}
\newcommand{\variablelogicspace}[2]{%
  \ifthenelse{\isempty{#2}}%
    {\ensuremath{#1}}
    {\ensuremath{{#1}{=}{#2}}}
  }
\newcommand{\vcut}[1][]{%
  \ifthenelse{\isempty{#1}}%
    {\ensuremath{V_{\rm cut}}}
    {\ensuremath{V_{\rm cut}{=}\kms[{#1}]}}
  }

\title[Astrophysical constraints on thermal relic DM]{Joint constraints on thermal relic dark matter from strong gravitational lensing, the Lyman-$\alpha$ forest, and Milky Way satellites}

\author[W. Enzi et al.]{
Wolfgang Enzi,$^{1}$\thanks{E-mail: enzi@mpa-garching.mpg.de }
Riccardo Murgia,$^{2}$
Oliver Newton,$^{3}$ 
\newauthor 
Simona Vegetti,$^{1}$
Carlos Frenk,$^{4}$
Matteo Viel,$^{5,6,7,8}$
Marius Cautun,$^{4,9}$
\newauthor 
Christopher D. Fassnacht,$^{10}$
Matt Auger,$^{11}$
Giulia Despali,$^{1,12}$
\newauthor
John McKean,$^{13,14}$
Léon V. E. Koopmans$^{13}$
and Mark Lovell$^{15}$
\\
$^{1}$Max Planck Institute for Astrophysics, Karl-Schwarzschild-Strasse 1, D-85740 Garching, Germany\\
$^{2}$Laboratoire Univers \& Particules de Montpellier (LUPM), CNRS \& Université de Montpellier (UMR-5299)\\
$^{3}$University of Lyon, UCB Lyon 1, CNRS/IN2P3, IUF, IP2I Lyon, France \\
$^{4}$Institute for Computational Cosmology, Department of Physics, Durham University, South Road, Durham, DH1 3LE, UK\\
$^{5}$SISSA-International School for Advanced Studies, Via Bonomea 265, 34136 Trieste, Italy\\
$^{6}$INFN, Sezione di Trieste, Via Valerio 2, I-34127 Trieste, Italy\\
$^{7}$INAF - Osservatorio Astronomico di Trieste, Via G. B. Tiepolo 11, I-34143 Trieste, Italy\\
$^{8}$IFPU, Institute for Fundamental Physics of the Universe, via Beirut 2, 34151 Trieste, Italy\\
$^{9}$Leiden Observatory, Leiden University, PO Box 9513, NL-2300 RA Leiden, the Netherlands\\
$^{10}$Department of Physics and Astronomy, University of California, Davis, USA\\
$^{11}$Institute of Astronomy, University of Cambridge, Madingley Road, Cambridge, CB3 0HA, UK\\
$^{12}$Zentrum für Astronomie der Universität Heidelberg, Institut für Theoretische Astrophysik, Albert-Ueberle-Str. 2, 69120 Heidelberg
 \\
$^{13}$Kapteyn Astronomical Institute, University of Groningen, P.O.Box 800, 9700AV, Groningen, the Netherlands\\
$^{14}$ASTRON, Netherlands Institute for Radio Astronomy, P.O. Box 2, 7990 AA Dwingeloo, the Netherlands\\
$^{15}$Center for Astrophysics and Cosmology, Science Institute, University of Iceland, Dunhagi 5, 107 Reykjavik, Iceland 
}

\date{Accepted XXX. Received YYY; in original form ZZZ}

\pubyear{2020}

\renewcommand{\P}{\mathcal{P}}

\defcitealias{newton_total_2018}{N18}

\defcitealias{newton2020constraints}{N20}

\defcitealias{vegetti2018}{V18}

\defcitealias{ritondale2019}{R19}

\defcitealias{hsueh2019}{H19}

\defcitealias{murgia2018}{M18}

\defcitealias{lovell_properties_2014}{L14}

\usepackage{xcolor}

\begin{document}
\label{firstpage}
\pagerange{\pageref{firstpage}--\pageref{lastpage}}
\maketitle

\begin{abstract}
We derive joint constraints on the warm dark matter (WDM) half-mode scale by combining the analyses  of a selection of astrophysical probes: strong gravitational lensing with extended sources, the Lyman-$\alpha$ forest, and the number of luminous satellites in the Milky Way. We derive an upper limit of $\lambda_{\rm hm}=0.089{\rm~Mpc~h^{-1} }$ at the 95 per cent confidence level, which we show to be stable for a broad range of prior choices. Assuming a Planck cosmology and that WDM particles are thermal relics, this corresponds to an upper limit on the half-mode mass of $M_{\rm hm }< 3 \times 10^{7} {\rm~M_{\odot}~h^{-1}}$, and a lower limit on the particle mass of $m_{\rm th }> 6.048 {\rm~keV}$, both at the 95 per cent confidence level.
We find that models with $\lambda_{\rm hm}> 0.223 {\rm~Mpc~h^{-1} }$ (corresponding to $m_{\rm th }> 2.552 {\rm~keV}$ and $M_{\rm hm }< 4.8 \times 10^{8} {\rm~M_{\odot}~h^{-1}}$) are ruled out with respect to the maximum likelihood model by a factor $\leq 1/20$.
 For lepton asymmetries $L_6>10$, we rule out the $7.1 {\rm~keV}$ sterile neutrino dark matter model,  which presents a possible explanation to the unidentified $3.55 {\rm~keV}$ line in the Milky Way and clusters of galaxies.
The inferred 95 percentiles suggest that we further rule out the ETHOS-4 model of self-interacting DM.
Our results highlight the importance of extending the current constraints to lower half-mode scales. We address important sources of systematic errors and provide  prospects for how  the constraints of these probes can be improved upon in the future.
\end{abstract}




\section{Introduction}

The nature of dark matter is one of the most important open questions in cosmology and astrophysics. While the standard cold dark matter (CDM) paradigm successfully explains observations of structures larger than $\sim 1$ {\rm Mpc}, it remains unclear whether observations on smaller (galactic and subgalactic) scales are consistent with this model \citep[e.g.][]{bullock2017}.
Possible alternatives include warm dark matter models \citep[WDM; e.g.][]{Bode_2001}, in which dark matter particles have higher velocities in the early Universe than in the CDM model. This characteristic leads to the suppression of 
gravitationally bound structures at scales proportional to the mean free path of the particles at the epoch of matter-radiation equality \citep[e.g.][]{schneider2012,lovell_properties_2014}. Until now, several complementary approaches have been used to test CDM and WDM on these scales. 
Among these are methods based on observations of strong gravitational lens systems, the Lyman-$\alpha$ forest, and the satellite galaxies of the Milky Way (MW).

Strong gravitational lensing, being sensitive only to gravity, allows one to detect low-mass haloes independently of their baryonic content. Therefore, it provides a direct method to quantify the dark matter distribution on subgalactic scales, where most of the structures are expected to be non-luminous. In practice, these low-mass haloes are detected via their effect on the flux ratios of multiply imaged compact sources \citep[flux-ratio anomalies;][]{Mao1998} or on the surface brightness distribution of magnified arcs and Einstein rings from lensed galaxies \citep[surface brightness anomalies or gravitational imaging;][]{koopmans2005, vegetti2009}. In this work, we focus on the latter method, while leaving the inclusion of analyses of flux ratios for future works. So far, both approaches have led to the detection of individual low-mass haloes \citep{vegetti2010a,vegetti2012, Nierenberg_2014,Hezaveh_2016_alma_detec}, as well as statistical constraints on the halo and subhalo mass functions, and on the related dark matter particle mass for sterile neutrino and thermal relic warm dark matter models \citep{dalal2002,vegetti2014,Birrer_2017,vegetti2018,ritondale2019,gilman2019_newest,hsueh2019}. In particular, the most recent analyses by \citet{hsueh2019} and \citet{gilman2019_newest} have derived a lower limit on the mass of a thermal relic dark matter particle of 5.6 and 5.2{\rm~keV} at the 95 per cent confidence level (c.l.), respectively.

While methods based on strong gravitational lensing target the detection of mostly dark low-mass haloes, the number of luminous satellite galaxies in the MW and other galaxies can also constrain the properties of dark matter
\citep[e.g.][]{moore_dark_1999,Nierenberg_2013}. For example, \citet{lovell_satellite_2016} compared the luminosity function of the MW satellites to predictions from semi-analytical models and derived lower constraints on the sterile neutrino particle mass of 2 {\rm keV}. 
More recently, by comparing the luminosity function of MW dwarf satellite galaxies to simulations and incorporating observational incompleteness in their model, \citet{Jethwa_2017} derived a lower limit of 2.9{\rm~keV} on the thermal relic particle mass at the 95 per cent confidence level.
\citet{Nadler_2019} derived a more stringent lower limit of ${\rm ~3.26~ keV}$ from the analysis of the classical MW satellites and those discovered by the Sloan Digital Sky Survey (SDSS).
 Combining data from the Dark Energy Survey \citep[DES, ][]{abbot2008} and the Pan-STARRS1 surveys \citep{chambers2019panstarrs1}, \citet{nadler2020milky}  derived a lower limit on the mass of thermal relic dark matter of  ${\rm ~6.5 ~keV}$ at the 95 per cent c.l. from the census of MW satellites.

The Lyman-$\alpha$ forest is one of the primary observational probes of the intergalactic medium \citep[IGM; see][for a review]{meiksin2009,McQuinn_2016}, and as such, it is used to probe the nature of dark matter as well as other cosmological quantities \citep{Narayanan2000,viel2005b,seljak2006,viel2006sndm,viel2008}. From the analysis of high-quality, high-resolution quasar absorption spectra at redshifts up to $z \approx 5.4$, \citet{irsic2017} constrained the lower limit of the thermal relic particle mass to be 5.3 {\rm~keV} at the 95 per cent c.l.  (3.5 {\rm~keV} with a more conservative prior, when assuming a smooth power-law and a free-form temperature evolution of the IGM). Recently, \citet{Murgia_2017,murgia2018} developed a broader approach to constrain generalised dark matter models \citep[e.g.][]{Archidiacono_2019,Miller_2019,baldes2020noncold,rogers2020general}, which in the case of thermal relic warm dark matter resulted in lower limits on the particle mass of 3.6 and  2.2 keV at the 95 per cent c.l. for the same assumptions on the thermal history of the IGM discussed above, respectively.

In this work, we extend and combine recent results from the three methods above \citep[][]{vegetti2018,murgia2018,ritondale2019,newton2020constraints} and derive joint constraints on the particle mass of a thermal relic dark matter model. This paper is structured as follows. We introduce the dark matter model under consideration in Section \ref{sec:dm_model_data}. In Section \ref{sec:methods_data}, we describe the method with which the different probes are analysed and combined. In Section \ref{sec:results}, we discuss the results obtained from the individual probes and their joint analysis. We discuss the different sources of systematic errors and the future prospects of each individual probe in Sections \ref{sec:systematics} and \ref{sec:future_prosepect}, respectively. Finally, we summarize the main results of this work in Section \ref{sec:conclusion}.

\section{Dark matter model}
\label{sec:dm_model_data}

We assume that dark matter is a thermal relic, that is, it consists of particles that were produced in thermodynamic equilibrium with photons and other relativistic particles in the early Universe\footnote{WDM-class particle models can exhibit very different production mechanisms and are not  necessarily in thermal equilibrium. For some of these models, e.g. sterile neutrino DM, the shapes of their linear matter power spectrum can still be well approximated by thermal relic models \citep[see][for an elaborate discussion]{lovell2020general}. This approximation is, however, not sufficient for all WDM-class models \citep[e.g.][]{dvorkin2020}.}. As the temperature of the Universe drops, dark matter decouples chemically and kinetically from the surrounding plasma (at the freeze-out time). Its density relative to the total entropy density of the Universe is then frozen in time \cite[see e.g. ][]{Bertone_2005} and it starts to free stream. As a result, the dark matter power spectrum is suppressed on scales related to the particles' free-streaming lengths and the size of the horizon at the time of decoupling. The warmer the dark matter particles (i.e., the larger their free-streaming length), the larger the scale at which the suppression happens. In this context, CDM and WDM belong to a continuum spectrum of free-streaming length or particle mass. From a statistical standpoint, this means that they are effectively nested models.

The cut-off in the WDM power spectrum $P_{\rm WDM}(k)$ can be expressed as a modification to the CDM power spectrum, $P_{\rm CDM}(k)$, via the following transfer function $T_{\rm matter}(k)$ \citep[see e.g.][]{Bode_2001}, i.e.:
\begin{equation}
T_{\rm matter}(k)^2 =   \frac{P_{\rm WDM}(k)}{P_{\rm CDM}(k)} = \Big( (1 + (\alpha k)^{2\mu_t} )^{-5/\mu_t} \Big)^2 \,,
\end{equation}
with the slope parameter $\mu_t = 1.12$ and the break scale $\alpha$, which for a given thermal relic density parameter $\Omega_{\rm th}$ and Hubble constant $h$ is determined to be \citep[see][]{viel2005}:
\begin{equation} 
\alpha   = 0.049 \Big(\frac{ m_{\rm th} }{1~{\rm keV} }\Big)^{-1.11 } \Big(\frac{ \Omega_{\rm th} }{0.25} \Big) ^{0.11} \Big(\frac{{\rm h}}{0.7} \Big)^{1.22}~{\rm Mpc~h^{-1}}.
\label{equ:alpha}
\end{equation}
{The half-mode scale\footnote{The definition of the half-mode scale differs within the literature and is sometimes defined so that $T^2(2 \pi / \lambda_{\rm hm})=1/2$. Here we follow the $T(2 \pi / \lambda_{\rm hm})=1/2$ convention.}, $\lambda_{\rm hm}$, at which the transfer function becomes equal to $1/2$, is then defined as:
\begin{equation}
 \lambda_{\rm hm} =  2 \pi \alpha \Big[ \big ( 0.5 \big)^{-\mu_t/5} -1 \Big]^{-\frac{1}{2\mu_t}}.
 \label{equ:hmscale}
\end{equation}}
The mass related to this length scale is referred to as the half-mode mass $M_{\rm hm}$, where $M_{\rm{hm}} = 0$ corresponds to the idealized CDM model (showing no cut-off) and $M_{\rm{hm}} \sim 10^{-6}~$M$_\odot$ is predicted for a CDM model of weakly interacting massive particles \citep[WIMPs;][]{Green2004,Schneider2013}. The suppression of the power spectrum manifests itself also in the mass function $\frac{d}{dm}n$, which describes how the (projected) number density of haloes $n$ changes as a function of the halo mass $m$. The suppression of low-mass haloes in WDM scenarios is well represented by a best fit multiplicative function of the form \citep{ schneider2012,lovell_properties_2014}:
\begin{equation}
\frac{d}{dm} n_{\rm WDM} = \frac{d}{dm} n_{\rm CDM}\left(1+\frac{M_{\rm hm}}{m}\right)^\beta\,,
\label{equ:mf_suppression}
\end{equation}
with a logarithmic slope $\beta \approx $ -1.3.
A more general parametrization that relates the CDM and WDM scenarios was recently developed by \citet{lovell2020general}. We leave the study of this more general parametrization for future works.
Combining equations (\ref{equ:alpha}) and (\ref{equ:hmscale})  the half-mode mass $M_{\rm hm} =  \bar \rho_{\rm m} \frac{4\pi}{3} (\lambda_{\rm hm}/2)^3$ and the thermal relic particle mass, $m_{\rm th}$, are related to each other according to:
\begin{equation}
m_{\rm th} = 2.32~ \Big( \frac{M_{\rm hm}}{10^9 M_\odot} \Big)^{-0.3} \Big( \frac{\Omega_{\rm th}}{0.25}  \Big)^{0.4}\Big(\frac{h}{0.7} \Big)^{0.8}~{\rm keV}\,.
\end{equation}

\section{Methods and Data}
\label{sec:methods_data}

In this section, we provide details of the data, models, and analyses that are the main focus of this paper. For each probe we rerun and/or extrapolate the analysis in order to match important model assumptions and to guarantee overlap in the prior range of $\lambda_{\rm hm}$.

\subsection{Strong gravitational lensing}
\label{sec:methods_data_lensing}

Galaxy-galaxy strong gravitational lensing occurs when the light from a background galaxy is deflected by the gravitational potential of another intervening galaxy. As a result, one observes multiple images of the background galaxy that are highly distorted and magnified. Substructures within the foreground lensing galaxy and low-mass haloes along the line of sight to the background object can produce additional perturbations to the lensed images, with a strength that depends on the mass of these (sub)haloes. Therefore, strong gravitational lensing provides a means to constrain the halo and subhalo mass functions directly.

\subsubsection{Halo and Subhalo model}

To describe the CDM field halo mass function we assume the formulation introduced by \citet{sheth1999}. For the subhalo mass function we assume a power-law,
\begin{equation}
 \frac{d}{dm}  n^{\rm sub}_{\rm CDM} = A \times m^{\-\gamma}\,,
\label{equ:mf}
\end{equation}
with a logarithmic slope $\gamma$ that is  -1.9 \citep{springel2008}.
In general, the amplitude of the subhalo mass function depends on the host redshift and mass, i.e. $A=A(M_{\rm vir},z)$ \citep{gao2011,han_unified_2016,chua2017}.
Here, we relate the normalisation constant $A$ to the average fraction of projected total host mass within two Einstein Radii which is contained in subhaloes, $f_{\rm sub}$. We discuss this assumption, i.e. $A = A(M(<2R_{\rm E}))$, in Section \ref{sec:systematics_lensing}. Substructures are assumed to be uniformly distributed within this area in agreement with the results of previous studies \citep{giulia_radial_dist,Xu_2015}. For a general dark matter model $f_{\rm sub}$ is equal to:
\begin{equation}
 f_{\rm sub} = \frac{ 4 \pi R_{\rm E}^2\int  dm~ m \times A \times m^{\-\gamma} \times  \Big ( 1+\frac{M_{\rm hm}}{m} \Big)^\beta }{M(<2 R_{\rm E}) } \,.
 \label{equ:fsub_mf_rel}
\end{equation}
 Solving the above equation for $A$ and setting $M_{\rm hm} = 0$, as is the case for ideal CDM, we find that the normalisation is determined to be:

\begin{equation}
A = \frac{M(<2 R_{\rm E})  f_{\rm sub}^{\rm CDM}}{ 4 \pi R_{\rm E}^2\int  dm~ m \times m^{\-\gamma} } \,.
 \label{equ:fsub_mf_rel2}
\end{equation}
Notice that while $A$ is independent of the DM model being warm or cold, according to equation (\ref{equ:fsub_mf_rel}), the value of $f_{\rm sub}$ in WDM models is related to its CDM counterpart according to:

\begin{equation}
 f_{\rm sub}^{\rm WDM}= f_{\rm sub}^{\rm CDM}\times \frac{\int  dm~ m \times m^{\-\gamma} \times  \Big ( 1+\frac{M_{\rm hm}}{m} \Big)^\beta }{\int  dm~ m \times m^{\-\gamma} } \,.
 \label{equ:fsub_change_wdm}
\end{equation}
The target parameters in our inference process are, therefore, $f_{\rm sub}^{\rm CDM}$  and $M_{\rm hm}$, since these parameters fully describe the mass function of subhaloes.
We assume that  $f_{\rm sub}^{\rm CDM} \in [0.01, 10]$ per cent with a uniform prior, which covers a wide range of previously  inferred values of $f_{\rm sub}^{\rm  CDM }$ with their uncertainties \citep[see e.g.][]{hsueh2019}.
Using this parameterization we can enforce that for each lens system the range of normalisations, $A$, is the same in WDM and CDM, and that the number of subhaloes scales with the projected mass of the lens galaxy. 
{The advantage of this approach compared to, e.g. \citet[][]{vegetti2018} and \citet[][]{ritondale2019}, is that it ensures that WDM models have fewer total substructures than in CDM as expected from theory.} 

Another difference to these previous works is our choice of the mass range for haloes.  We chose a mass range of subhaloes of $m_{\rm sub} \in [10^6,10^9]M_{\odot} {~\rm h}^{-1}$. The upper limit is chosen so that this range includes only masses corresponding to objects that are not expected to be visible, because they are  either non-luminous or too faint to be observed \citep[see e.g.][]{Moster_2010}. 
We choose the mass range of line-of-sight haloes such that it contains the masses that show the most similar lensing effects to the lightest and heaviest substructures according to the mass-redshift relationship derived by \citet{despali2018}, $m_{\rm los} \in [10^{5.26},10^{10.88}]M_{\odot}{~\rm h}^{-1}$.

\begin{table}
	\centering
	\caption{Main model parameters constrained by strong  lensing observations and their relative prior ranges. From top to bottom: the virial mass of subhaloes and field haloes, the half-mode mass, the fraction of mass in subhaloes 
	(note that $f_{\rm sub}$ is defined differently in the original analyses of \citetalias{vegetti2018} and \citetalias{ritondale2019}).
	}	
	\label{tab:params_lensing}
	\begin{tabular}{lcr}
	\hline
	Parameter & Original & This paper\\
	\hline
	$m_{\rm sub} \left[M_{\odot}{~\rm h}^{-1}\right]$ & $[ \approx 10^{5} , 2 \times 10^{11}]$& $[10^{6},10^{9}]$
	\\
	$m_{\rm los} \left[M_{\odot}{~\rm h}^{-1}\right]$ & $[ \approx 10^{5} , 2 \times 10^{11}]$& $[10^{5.26}, 10^{10.88}]$
	\\
	$M_{\rm hm}\left[M_{\odot}{~\rm h}^{-1}\right]$ & $[10^6,2\times 10^{12}]$ & $[10^{-6}, 10^{14}]$
	\\
	$f_{\rm sub}\left[\%\right]$& [0.0, 4.0]&  $[0.01, 10.0]$\\
	\hline
\end{tabular}
\end{table}

For both populations of haloes, the suppression in the number density at the low-mass end is calculated using equation (\ref{equ:mf_suppression}).
While this suppression relative to the CDM case tends to be stronger in the case of field haloes  than in subhaloes, we ignore this effect in this work for simplicity 
\citep{lovell2020general}.
 Both halo populations are assumed to have spherical NFW profiles \citep{navarro1996} following the mass-concentration-redshift relation derived by \citet{duffy2008}.  We discuss these assumptions in Section \ref{sec:systematics_lensing}. In Table \ref{tab:params_lensing}, we summarize all of the relevant parameters, together with the corresponding priors used in this paper and previous analyses. 
For the lensing analyses, we adopt the cosmology inferred by the Planck mission \citep{Planck2013}.

\subsubsection{Data}

We consider the re-analyses of \citet{vegetti2010a,vegetti2014}, who analysed a subsample of eleven gravitational lens systems from the SLACS survey \citep{bolton2008}.  Using the Bayesian gravitational imaging technique developed by \citet{vegetti2009}, only one low-mass subhalo was detected in the sample. Assuming a Pseudo-Jaffe profile, this subhalo shows an inferred mass of $3.5\times10^9 M_{\odot}$ ($\sim 10^{10} M_{\odot}$ for an NFW profile). Taking this detection and the non-detections into account, they constrained the subhalo mass function to be consistent with CDM.
The lenses in this sample have a mean redshift of $z=0.2$, while the background sources have a mean redshift of $z=0.6$. 
In the remaining part of this paper, we refer to this sample as the low-redshift sample.

The background source galaxies were modelled in a free-form fashion with a Delaunay mesh, while the foreground lenses were assumed to have an elliptical power-law mass density profile plus the contribution of an external shear component. Additional forms of complexity in the lenses not captured by the smooth power-law (including subhaloes) were identified using linear free-form corrections to the lensing potential. The statistical relevance of both detections and non-detections is determined via the sensitivity function.  This function considers the Bayes factor between models with no substructure and those with a single substructure. A logarithmic Bayes factor of 50 provided a robust criterion to discriminate between reliable and non-reliable detections.  Originally this description assumed a Pseudo-Jaffe parametric profile for the perturber \citep[][]{vegetti2014}.  The analysis by \citet{vegetti2014} was then extended by  \citet[][hereafter \citetalias{vegetti2018}]{vegetti2018}. This new analysis includes the contribution of low-mass field haloes \citep[i.e. haloes located along the line of sight; see][]{despali2018,Li2017}, and changed the density profile of subhaloes from a Pseudo-Jaffe to an NFW profile. Furthermore, the effects of dark matter free streaming on the halo and subhalo mass functions were included via equation (\ref{equ:mf_suppression}). 

\citet[][hereafter \citetalias{ritondale2019} ]{ritondale2019} have modelled a sample of seventeen gravitational lens systems from the BELLS-GALLERY survey \citep{shu2016}, and reported zero detections of subhaloes and line-of-sight haloes. The mean redshift of the foreground lenses is $z\sim0.5$, while the background source redshifts vary from $z=2.1$ to $2.8$. We refer to this sample as the high-redshift sample. The analysis by \citetalias{ritondale2019} used a more recent version of the \citet{vegetti2009} lens modelling code that allows for a simultaneous inference on the lens galaxy mass and light distribution. As in the original method, the source surface brightness distribution and low-mass haloes are defined on a grid of pixels. The calculation of the sensitivity function and the inference on the mass functions were performed in terms of the spherical NFW virial mass.

Here, we re-run the analyses of \citetalias{vegetti2018} and \citetalias{ritondale2019}, while extending their prior ranges on the half-mode mass to $M_{\rm hm} \in [10^{-6},10^{14}]~{\rm ~M_{\odot}~h^{-1}}$.

\subsection{\texorpdfstring{Lyman-$\alpha$}{Lyman-alpha} forest}

The second astrophysical probe that we consider comes from the analysis of high-quality optical spectra from Lyman-$\alpha$ emitting quasars at high redshifts. As the quasar light travels through the Universe, the spectrum becomes correlated with the matter power spectrum at different redshifts. In particular, the quasar light is redshifted during the expansion of the Universe, so that Lyman-$\alpha$ absorption from neutral hydrogen clouds along the line of sight suppresses different parts of the original quasar spectrum at each redshift. As the matter power spectrum depends on the dark matter model, the comparison between mock spectra obtained from hydrodynamical simulations and those retrieved from spectroscopic observations can constrain the nature of dark matter. 

The approach of \citet[hereafter \citetalias{murgia2018}]{murgia2018} obtains constraints on dark matter models by performing a Monte Carlo Markov Chain (MCMC) analysis of the full parameter space affecting the flux power spectrum $P_{\rm f}(k)$ reconstructed from high-redshift Lyman-$\alpha$ forest observations.
We rerun the analysis of \citetalias{murgia2018} changing two main elements; first, the results presented here are restricted to the analysis of thermal relic warm dark matter, for which we choose a log-uniform prior on the particle mass, $m_{\rm WDM}$. Second, we extend the ranges of some model parameters, which are discussed below.

Due to our focus on thermal relic warm dark matter, we do not fully take advantage of the versatility of the parametrization introduced by \citet{Murgia_2017}. We refer the reader to \citetalias{murgia2018} and \citet{rogers2020general} for a demonstration of the complete flexibility of this parametrization in the study of different non-thermal dark matter models.

The data set used for the analysis is provided by the high-resolution and high-redshift quasar samples from the HIRES/Keck and the MIKE/Magellan spectrographs \citep{viel2013}. These samples include redshift bins of $z = 4.2$, 4.6, 5.0 and 5.4 over 10 wavenumber bins in the interval $k\in  [0.001,0.08]~{\rm s~ km^{-1}}$ (the range relevant for Lyman-$\alpha$ forest data). The spectral resolution of the HIRES and MIKE spectrographic data are 6.7 and $13.6 {\rm~ km~ s^{-1}}$,  respectively.
As in previous analyses, such as \cite{viel2013}, only the measurements with $k > 0.005~{\rm s ~km^{-1}}$ have been used to avoid systematic uncertainties on large scales due to continuum fitting. The highest redshift bin for the MIKE data has been excluded due to the large uncertainties in the spectra at that epoch (see \citealt{viel2013} for further details). A total of 49 data points in wavenumber $k$ and redshift $z$ are used in the analysis.

\citetalias{murgia2018} determined the changes in the flux power spectra as a function of different model parameters by interpolation of (computationally expensive) realistically simulated mock spectra, which are generated for different astrophysical and cosmological parameters defined on a grid. This procedure allowed \citetalias{murgia2018} to define a likelihood as a function of these parameters.
The grid of mock simulations considers several values of the cosmological parameters and follows the approach of \citet[][]{irsic2017} to recover their effects on the likelihood.  \citetalias{murgia2018} considered five different values for the normalisation of the linear matter power spectrum $\sigma_8 \in  [0.754, 0.904]$ and its slope  $n_{\rm eff} \in [-2.3474,-2.2674]$ (both defined on the typical scale probed by the Lyman-$\alpha$ forest of $0.009$~s\,km$^{-1}$). For the rerun of this analysis, we consider ten~$\Lambda$WDM simulations that correspond to thermal WDM masses of $m_{\rm wdm}\in[1,10]${\rm~keV}, linearly spaced in steps of 1{\rm~keV}.

Concerning the astrophysical parameters, the thermal history of the IGM is varied in the form of the amplitude $T_0$ and the slope $\gamma$ of its temperature-density relation. This relation is parametrized as $T=T_0(1+\delta_\mathrm{IGM})^{\gamma-1}$, with $\delta_{\mathrm{IGM}}$ being the overdensity of the IGM \citep{hui1997}. Three different temperatures at mean density, $T_0(z = 4.2) = 6000$, 9200 and 12600~K, and three values for the slope of the temperature-density relation, ${\gamma}(z = 4.2) = 0.88$, 1.24 and 1.47, are considered here. The reference thermal history is defined by $T_0(z = 4.2) = 9200$~K and ${\gamma}(z = 4.2) = 1.47$ \citep[see][]{bolton2016}. The redshift evolution of $\gamma(z)$ is assumed to be a power law, that is, $\gamma(z) = {\gamma}^A[(1+z)/(1+z_p)]^{{\gamma}^S}$, where the pivot redshift $z_p$ is the redshift at which most of the Lyman-$\alpha$ forest pixels originate ($z_p = 4.5$ for the MIKE and HIRES datasets).  

\citetalias{murgia2018} also considered three different redshift values of instantaneous reionization at $z_{\rm reion} \in \{ 7,9,15\}$, as well as ultraviolet (UV) fluctuations of the ionizing background of $f_{\rm UV} \in \{ 0, 0.5, 1\}$, where the case of $f_{\rm UV} = 0$ corresponds to a spatially uniform UV background. Nine values of the relative mean flux were considered, that is, $ \langle F(z) \rangle / \langle F_{\rm REF} \rangle \in [0.6,1.4]$ in linearly spaced intervals with steps of $0.1$. The reference values $\langle F_{\rm REF}\rangle$ are taken from the Sloan Digital Sky Survey (SDSS), i.e. the Baryon Oscillation
Spectroscopic Survey (BOSS) measurements, which are part of SDSS-III \citep{anderson2014}. Eight additional values of $\langle F(z)\rangle / \langle F_{\rm REF}\rangle$ are obtained by rescaling the optical depth $\tau = -{\ln \langle F \rangle}$ \citepalias[see ][]{murgia2018}.

For each of the resulting grid points in parameter space, hydrodynamical simulations are used to generate the mock spectra. All simulations are performed with {\tt GADGET-3}, a modified version of the publicly available {\tt GADGET-2} code \citep{springel2005,springel2001}. As in \cite{irsic2017}, 
the reference model simulation has a box length of 20~${\rm Mpc~h^{-1}}$ (comoving) with $2 \times 768^3$ gas and CDM particles (with gravitational softening lengths of 1.04~${\rm kpc~h^{-1}}$ comoving) in a flat $\Lambda$CDM Universe. The cosmological parameters are set to $\Omega_{\rm m}= 0.301$, $\Omega_{\rm b} = 0.0457$, $n_s = 0.961$, $H_0 = 70.2$~${\rm km~s^{-1}~Mpc^{-1}}$ and $\sigma_8 = 0.829$ \citep[][]{planck2016}.

An Ordinary-Kriging scheme is used for the interpolation between different grid points and linearly extrapolated when necessary \citep{webster2007geostatistics}.
The interpolation with respect to all the parameters happens in consecutive steps, first over the astrophysical and cosmological parameters, then over the different WDM models. This interpolation is then used to define a likelihood, which in return produces a posterior \citep[e.g.][]{Archidiacono_2019}. Table \ref{tab:params_lensing_LY} gives a short overview of the model parameters, their ranges and (prior) probabilities. We replace the original prior on the WDM particle mass with a log-uniform prior in order to match the priors of the other probes considered here. 
We note that MCMC analyses of the Lyman-alpha forest would be computationally infeasible without a fast and efficient interpolation scheme. The main reason is that each MCMC step would require the output of a hydrodynamical simulation when exploring the parameter space. A possible alternative to the Ordinary-Kriging interpolation is a Bayesian optimization emulator \citep[see e.g][]{rogers2020general,pedersen2020}.

\begin{table}
	\centering
	\caption{
	The model parameters, their ranges, and their (prior) probability distributions, as they are used in the rerun of the analysis in \citetalias{murgia2018}. \\${}^{(*)}$Is the same prior as described in \citet{irsic2017}.
	}
	\label{tab:params_lensing_LY}
	\begin{tabular}{lcr} 
		\hline
		Parameter &   Range &  Probability\\
		\hline
$1/m_{ \rm wdm}$ $[{\rm keV^{-1}}]$ & [0, 1] & Log-flat \\
$\langle F(z) \rangle/ \langle F_{\rm REF} \rangle$ & $(-\infty,\infty)$ &  Gaussian$^{(*)}$
\\
$T_0(z)$ $[10^4 {\rm K}]$ &$[0, 2]$ & Flat
\\
$\widetilde{\gamma}(z)$  & [1, 1.7] & Flat
\\
$\sigma_8$   & [0.5, 1.5] & Flat
\\
$z_{\rm reion}$  & [7, 15] & Flat
\\
$n_{\rm eff}$   & [-2.6,-2.0]& Flat
\\
$f_{\rm UV}$   & [0,1] & Flat \\
		\hline
	\end{tabular}
\end{table}

\subsection{Milky Way luminous satellites}
\label{sec:milky_way}

Our final astrophysical probe comes from the observed luminosity function of the satellite galaxies of the MW. This method was the first to identify a potential challenge to the CDM model due to the paucity of observed dwarf galaxies around our own galaxy \citep{kauffmann1993,klypin_where_1999,moore_dark_1999}.
Observational solutions, such as the lack of sufficient sky coverage/completeness \citep{koposov2008,TOllerud2008,Hargis_2014,Jethwa_2017,Kim_2018}, more realistic galaxy formation models  \citep[e.g. stellar feedback,][]{bullock_reionization_2000,benson_effects_2002,somerville_can_2002,burkert2004,Agertz_2016,De_Lucia_2018} or revisions to the dark matter model \citep{Bode_2001,Green2004,schneider2012} have since been proposed to reconcile this discrepancy. Here, we consider a new analysis of the number density of luminous satellites that has been carried out by  \citet[][\citetalias{newton2020constraints}, hereafter]{newton2020constraints}. 

Their approach assesses the viability of a given WDM model by comparing the predictions of the abundance of satellite galaxies within a MW-mass halo for various dark matter models with the total satellite galaxy population inferred from those observed in the MW.
WDM scenarios that do not produce enough faint galaxies to be consistent with the MW satellite population are ruled out with high confidence. As the current census of MW satellites is incomplete, \citetalias{newton2020constraints} infer the total satellite galaxy population from observations, using a Bayesian formalism that was developed and tested robustly by \citet[][]{newton_total_2018}. 
They use data from the SDSS and DES, as summarized in Table A1 of \citetalias[][]{newton_total_2018} \citep[compiled from][]{watkins_substructure_2009,mcconnachie_observed_2012,drlica-wagner_eight_2015,kim_heros_2015,koposov_kinematics_2015,jethwa_magellanic_2016,kim_portrait_2016,walker_magellan/m2fs_2016,carlin_deep_2017,li_farthest_2017}. More recent discoveries of dwarf galaxy candidates are not incorporated into our analysis. However, it is unlikely that their inclusion would change the inferred population outside the uncertainties quoted in \citetalias{newton2020constraints}. 

For each survey, the assumed observational selection function significantly affects the size of the total satellite population inferred from the observations. In particular, if the selection function overpredicts the completeness of faint objects in the survey, then the size of the inferred satellite population will be too small. While the SDSS selection function has been studied extensively and is now well-characterized \citep[e.g.][]{walsh_invisibles_2009}, no such study had been carried out for the DES before 2019. Therefore, \citetalias{newton2020constraints} 
used the approximation proposed by \citet{jethwa_magellanic_2016}, one of the few estimates available for the DES at the time. The DES selection function was recently characterized in detail by \citet{drlica-wagner_milky_2019} and used in follow-up studies by \citet{nadler_milky_2019, nadler2020milky} to infer the total satellite population. Their results are consistent with \citet{newton_total_2018} and \citetalias{newton2020constraints}, 
which suggests that the \citet{jethwa_magellanic_2016} approximation of the selection function was reasonable.

The second ingredient of the analysis by  \citetalias{newton2020constraints}  is a set of estimates of the number of satellite galaxies formed in MW-mass WDM haloes.
They explore two approaches to obtain these predictions. In the first, they use the Extended Press-Schechter~(EPS) formalism \citep{press_formation_1974,bond_excursion_1991,bower_evolution_1991,lacey_merger_1993,parkinson_generating_2008} and follow the approach of \citet{kennedy_constraining_2014}, \citet{schneider_structure_2015}, and \citet{lovell_satellite_2016}. Implicit in this technique is the assumption that all DM haloes form a galaxy, which allows \citetalias{newton2020constraints} to place a highly robust lower bound on the mass of the warm dark matter particle independently of assumptions about galaxy formation physics. However, the faint end of the satellite galaxy luminosity function is extremely sensitive to these processes, which prevent galaxies from forming in low-mass DM haloes.
They are also complex and their details remain uncertain, permitting a large parameter space of viable descriptions of galaxy formation.
In their second approach,  \citetalias{newton2020constraints}  use the \Galform{} semi-analytic model of galaxy formation \citep{cole_recipe_1994,cole_hierarchical_2000} to explore this space to understand how different parametrizations can affect the WDM constraints.
The main process affecting the MW satellite galaxy luminosity function is the reionization of the Universe. In \Galform{}, this is described by $z_{\rm reion}$, the redshift at which the intergalactic medium is fully ionized, after which the parameter $V_{\rm cut}$ prevents cooling of gas into haloes with circular velocities, ${v_{\rm vir} <V_{\rm cut} }$.  \citetalias{newton2020constraints} assume the \citet{lacey_unified_2016}  version of galaxy formation and vary the reionization parameters in the ranges $6 \leq z_{\rm reion} \leq 8$ and $25 \kms \leq V_{\rm cut} \leq 35 \kms$, choosing a fiducial model with $z_{\rm reion}=7$ and $V_{\rm cut} =30 \kms$. From each \Galform{} model they obtain MW satellite galaxy luminosity functions which they compare with the luminosity functions inferred from observations (described above). They calculate the relative likelihood of a given model compared to the CDM case by convolving the probability density function of the number of satellites brighter than $M_{\rm V} =0$ in a \Galform{} WDM MW halo with the cumulative distribution function of the inferred population of MW satellites, which, according to \citet[][]{newton_total_2018}, numbers $124^{+40}_{-27}$. We use this  approach in the analysis that follows.

Comparing this approach to the recent study by \citet{nadler2021}, there are two important differences. First, the results by \citet{nadler2021} are based on data from DES + Pan-STARRS rather than DES + SDSS, which we use in this work. This difference leads to an inferred number of MW satellites which is roughly twice as large as the one by \citetalias[][]{newton2020constraints}.  When applied to SDSS data alone, both methods estimate roughly the same number of MW satellites \citep[see Fig. 8 in][]{nadler2019_modeling}, which suggests that the discrepancy is due to differences in the detection efficiency of satellites  in SDSS and Pan-STARRS \citep{drlica-wagner_milky_2019}. 
A second difference between the two approaches is that \citetalias[][]{newton2020constraints} used the {\sc Galform} semi-analytic galaxy formation model, while \citet{nadler2021} used a halo occupation distribution model. It remains to be studied what the exact implications of these choices are. However, the fact that the two methods obtain consistent results when applied to the same SDSS data, already suggests that the specific choice of model may not be critical.

\subsection{Model consistency}
\label{sec:homogeneity} 

The goal of this paper is to derive joint constraints on the particle mass of a thermal relic dark matter model. However, $m_{\rm th}$ is not directly observable but is inferred from the different probes under different assumptions, as described above.
In this section, we discuss the main differences between the three methods and how these can be treated to derive a meaningful joint inference on $m_{\rm th}$.

All the methods employed here constrain parameters describing the halo and mass function; however, these parameters may differ in their meaning. The main parameter constrained by strong gravitational lensing observations is the half-mode mass $M_{\rm hm}$, while the analysis of the Lyman-$\alpha$ forest and the luminosity function of the MW satellites are expressed directly in terms of $m_{\rm th}$. Converting from one to the other requires some assumptions about the physics of the dark matter particles (e.g. their type and production mechanism; see Section \ref{sec:dm_model_data}) and the cosmological parameters. As each of the considered analyses has adopted different cosmologies, we first express our inference in terms of the half-mode scale $\lambda_{\rm hm}$, which is less sensitive to the specific values of the cosmological parameters.

For all the different probes, we assume a uniform prior in $\lambda_{\rm hm}$ with the lowest possible value corresponding to the WIMP CDM model, that is, $M_{\rm hm}(\lambda_{\rm hm}^{\rm min}) = 10^{-6}$ ${\rm~M_{\odot}~ h^{-1}} $, and the upper limit $M_{\rm hm}(\lambda_{\rm hm}^{\rm max}) = 10^{14}$~${\rm~M_{\odot} ~h^{-1} }$, which corresponds to the lower limit on a thermal relic WDM particle mass $m_{\rm th} = 0.07 {\rm~ keV}$ as constrained by \citet{kunz2016} using observations of the cosmic microwave background \citep{planck2016b}. 
We then express our results in terms of the half-mode and thermal relic particle masses, converting all results so that they adopt Planck cosmology \citep[i.e. $\Omega_{\rm th} = 0.26$ and ${\rm h}= 0.68$;][]{planck2016} and the assumptions on the dark matter particles described in Section \ref{sec:dm_model_data}. Notice that logarithmic quantities $\log_{10}(M_{\rm hm})$ and $\log_{10}(m_{\rm th})$ are related to  $\log_{10}(\lambda_{\rm hm})$ via linear transformations, so that the prior is flat in all of these parameters.

Another potential difference could arise from the models used to describe the population of low-mass haloes. 
The lensing analyses make direct use of the halo and subhalo mass function (see Section \ref{sec:methods_data_lensing}) expressed in terms of the virial mass of a spherical NFW profile. The analysis of the MW satellites depends only on the radial distribution of the subhaloes, independent of their present-day mass or the details of their profile \citepalias{newton2020constraints}. 
The Lyman-$\alpha$ forest constraints are expressed in terms of the matter power spectrum.
\citet{despali2018} have shown that the lensing effect of an NFW subhalo of a given mass $M_{\rm vir}^{\rm NFW}$ is a good approximation to a subhalo of equivalent mass found with the SUBFIND algorithm \citep{Springel_2001} in cosmological simulations. This indicates that the lensing treatment of the subhalo masses is consistent with the adopted parametrization of the subhalo mass function. 
Moreover, as all probes have been calibrated on numerical simulations, we can assume that there is no strong discrepancy between the use of mass functions, number counts or power spectra. Given these considerations, we conclude that any discrepancies in the treatment of low-mass haloes are negligible and can be ignored.

\section{Results}
\label{sec:results}
 
In this section, we present our main results on the half-mode scale and mass, and the thermal relic particle mass. We present the constraints from each of the individual probes as well as those of the joint statistical analysis. Our statistical summaries are presented in Table \ref{tab:posterior_summaries}, which can be compared with the previous results that we report in Table~\ref{tab:comp}.

\subsection{Posterior distributions}

\begin{table*}
\centering
\caption{The posterior limits according to the 95 percentile criterium (Section \ref{sec:statsum_95}) and the Bayes factor (BF, Section \ref{sec:statsum_BF}), probabilities of (in-)sensitivity, odds of being sensitive, ratio between the likelihoods of the maximum a likelihood half-mode mass $\lambda^{\rm ML}_{\rm hm}$ and that of the insensitive region.
}
\label{tab:posterior_summaries}
\begin{tabular}{lccccccccccc}
\hline
Reference &\multicolumn{2}{|c|}{ $\frac{m_{\rm th}}{\rm keV}$} & \multicolumn{2}{|c|}{ $ \frac{M_{\rm hm}}{ 10^{10} {\rm ~M_{\odot}~  h^{-1}} }$} 
& \multicolumn{2}{|c|}{ $ \frac{\lambda_{\rm hm}}{{\rm Mpc~h^{-1}}}$} & $\P(\bar S| d) $& $\P( S |  d)$ & $\frac{\P(\bar S| d)}{\P(S| d)}$ & $\frac{\P(d |\lambda_{\rm hm}^{\rm ML})}{\P( d |\lambda_{\rm hm}\in \bar S)}$  & $\frac{\lambda_{\rm hm}^{\rm ML}}{{\rm Mpc~h^{-1}}}$\\
 & BF  &$95\%$ c.l. & BF  &$95\%$ c.l. & BF  &$95\%$ c.l.  & \% & \% &1 & 1 & 1 \\
 \hline
 \citetalias{vegetti2018} &
0.216 & 0.576 & 178.366 & 6.780 & 3.607 & 1.214 & 47.52 & 52.48 & 0.91 & 3.35 & 0.470 \\
  \citetalias{ritondale2019} &
- & 0.121 & - & 1219.752 & - & 6.842 & 53.06 & 46.94 & 1.13 & 1.14 & 4.538 \\
 \citetalias{murgia2018}  & 
 1.197 & 3.571 & 0.594 & 0.016 & 0.540 & 0.160 & 74.30 & 25.70 & 2.89 & 1.04 & 0.029 \\
\citetalias[][]{newton2020constraints} & 
2.678 & 6.989 & 0.041 & 0.002 & 0.221 & 0.076 & 79.46 & 20.54 & 3.87 & 1.01 & 0.016 \\
\hline
Joint & 
2.552 & 6.048 & 0.048 & 0.003 & 0.233 & 0.089 & 77.68 & 22.32 & 3.48 & 1.08 & 0.027 \\
\hline
\end{tabular}
\end{table*}

\begin{table}
\centering
\caption{
A summary of the lower limits reported on the thermal relic dark matter particle mass for a selection of past studies. Note that additional model assumptions and assumed parameter ranges can widely differ. When derived for different assumptions, we provide more than one of the limits. }
\label{tab:comp}
\begin{tabular}{lcr} 
\hline
Reference & Probe&  $ \frac{m_{\rm th} }{\rm keV} $  \\
& &  $95\%$ c.l.\\
\hline
this work &  see Section \ref{sec:methods_data} & 6.048
\\
\hline
\citet{Birrer_2017} & Grav. Imaging &  2.0  
\\
\citetalias{vegetti2018} (Original) & Grav. Imaging  &  0.3 
\\
\citetalias{ritondale2019} (Original)  &Grav. Imaging &  0.26
\\
\citet{gilman2019}   & Flux Ratios&  3.1, 4.4  \\
\citet{gilman2019_newest}   &Flux Ratios  & 5.2 
\\
\citet{hsueh2019}  & Flux Ratios & 5.6
\\
\citet{Banik_2018_naturedm,banik2019novel} & Stellar streams 
& 4.6, 6.3  \\ 
\citet{alvey2020new}& Dwarf spheroidals & 0.59, 0.41 \\
\citet{viel2005b}   & Lyman-$\alpha$  &0.55 
\\
\citet{viel2006sndm}    & Lyman-$\alpha$&  2.0 	 
\\
\citet{seljak2006}  & Lyman-$\alpha$ & 2.5 \\
\citet{irsic2017}  &  Lyman-$\alpha$ &  3.5, 5.3  
\\
\citetalias{murgia2018} (Original) & Lyman-$\alpha$ & 2.7, 3.6 
\\
\citet{Polisensky_2011} & MW satellites &   2.3   
\\
\citet{kennedy_constraining_2014}  & MW satellites &  1.3, 5.0  
\\
\citet{Jethwa_2017} & MW satellites & 2.9  \\
\citet{Nadler_2019}& MW satellites &  3.26\\
\citet{nadler2020milky}& MW satellites &  6.5\\
\citet{nadler2021}& MW satellites  & 9.7 \\
&\& Flux Ratios& \\
 \citetalias{newton2020constraints} (Original) & MW satellites &  2.02, 3.99\\
\hline
\end{tabular}
\end{table}

Fig.~\ref{fig:thermal_part_mass} shows the individual and the joint posterior on the half-mode scale, half-mode mass and thermal relic particle mass for each of the astrophysical probes considered here. Each of the posteriors is scaled so that its maximum value is equal to 1.

The shape of the joint posterior is mostly determined by the posterior of the analysis of luminous satellites in the MW galaxy \citepalias{newton2020constraints}, which is roughly shaped like a sigmoid function in a region of parameter space where the other posteriors are approximately flat. 
As a result, the joint posterior provides constraints that are close to but slightly weaker than the stand-alone analysis by \citetalias{newton2020constraints}.
The Lyman-$\alpha$ forest \citepalias{murgia2018} analysis - although being a completely independent measurement - finds a slightly higher upper limit on the half-mode scale. Further data and more rigorous analyses may reveal larger differences between their respective constraining power in the future.  

{
As a result of their weak constraints, neither of the lensing analyses contribute significantly to  the joint posterior.
Since the analysis of \citetalias{vegetti2018} includes the detection of a relatively massive subhalo \citep{vegetti2010a}, it only rules out higher values of $\lambda_{\rm hm}$  (as well as $M_{\rm hm} $) than the other probes. 
In contrast, the constraints from the BELLS-GALLERY sample turn out to be rather weak. As \citetalias{ritondale2019} reported no significant detections, the resulting posterior slightly prefers warmer dark matter models that predict a smaller number of (sub-) haloes. This may also be related to the sensitivities and the source redshifts of these lenses. 
While the higher sensitivity of the SLACS sample allows us to detect objects with smaller masses, the high redshift sources of the BELLS-GALLERY systems probe a larger cosmological volume increasing the expected number of line-of-sight objects and the statistical significance of the non-detections.

}

\subsection{Marginal vs non-marginal posteriors}
\label{sec:marginal}

In this paper, we have derived constraints on the half-mode scale by combining the individual posterior distributions marginalised over the nuisance parameters. We notice that joining the multidimensional posteriors before marginalizing may in theory lead to the breaking of degeneracies and, therefore, improved constraints.   
In practice, for this approach to work, one either needs nuisance parameters which are common to the different probes or a way to robustly connect different parameters.
For example, in their recent study, \citet{nadler2021} assumed a linear relation between the normalization of the subhalo mass function of the MW and typical lensing galaxies to account for the difference in the host halo masses, redshifts and morphologies. They concluded that this approach improves their constraints by as much as $\sim30$ per cent.

However, how the number of subhaloes depends exactly on the host galaxy properties is still poorly constrained.
N-body simulations show that the amount of subhaloes increases with the host mass and redshift \citep[see e.g.][]{Xu_2015,Gao_2010} 
and that it is set by a combination of the host accretion history and the number of subhaloes that survive tidal disruption processes. Unfortunately, these processes are difficult to model accurately in their full complexity. As recently shown by \citet{Green_2021} numerical artifacts in simulations may lead to an artificial disruption of subhaloes which can be as large as 20 per cent. This effect is problematic also for semi-analytical models which are traditionally calibrated on numerical simulations. We also know that the presence of baryons leads to a further disruption, which depends on the host morphology as well as the exact implementation of feedback processes 
and how they affect the host and the subhalo mass density profile \citep[][]{sawala_shaken_2017,garrison2017,giulia_radial_dist}.
Furthermore, all these effects are sensitive to the physics of dark matter in a way that has not yet been systematically quantified. Given these uncertainties,  we conclude that, for the analyses considered in this paper, joining the marginalised posterior distribution is expected to be less precise but probably more accurate.

\subsection{Statistical Summaries}

It is a common practice to report summary statistics of the posterior functions to characterise the strength of constraints on warm dark matter. One of the most reported quantities is the 95 percentile.
However this comes with a caveat:
the values of percentiles are strongly dependent on the specific choice of the lower limit of the model parameter range,
since the likelihood (and posterior) functions become essentially flat for $\lambda_{\rm hm} < 0.013 { \rm~Mpc~h^{-1}} $ (corresponding to $M_{\rm hm}< 10^{5.0}  {\rm~M_{\odot} ~ h^{-1}} $).
This flattening reflects a lack of sensitivity on these scales, i.e. that the analyses considered in this work are no longer capable of distinguishing between models of different half-mode scales. 

In the posteriors shown in Fig.~\ref{fig:thermal_part_mass}, we choose a lower-limit of $\lambda_{\rm hm} = 3 \times 10^{-6} {\rm ~ Mpc~ h^{-1}}$ which corresponds to a WIMP CDM model \citep{schneider_halo_2013}. We chose this limit mainly because a log-uniform prior gives rise to a diverging posterior if we extend the inference to the idealised CDM case of $\lambda_{\rm hm} = 0$. However, it could be argued that even though our choice of lower limit in the parameter range is physically motivated, it arbitrarily excludes models that lie between the WIMP and the idealised case. 

To account for some of the uncertainties in these a priori choices, we report two statistical summaries: one equivalent to the 95 percentiles within a rephrased version of the inference problem; the other based on the ratio of likelihoods and therefore, more independent of the chosen lower limit for $\lambda_{\rm hm}$ (and its prior).  Notice that this does not affect our main conclusions, but only accommodates for different preferences in the way that posteriors are summarized.

\subsubsection{95 percentiles}

 \label{sec:statsum_95}

For the first summary, we rephrase our inference problem in terms of a hyper-model scenario with two models corresponding to an insensitive ($\bar S$) and a sensitive ($S$) region, respectively. In particular, we define the former as the range of half-mode masses $M_{\rm hm} \in \left[0 , 10^{5}\right]{\rm M_{\odot}~ h^{-1}}$ and the latter as $ M_{\rm hm}   \in\left[10^5, 10^{12}\right]{\rm M_{\odot} ~ h^{-1} }$. We know that the likelihood in the two regions is then defined as follows:
     \begin{multline}
    \P\Big(d\Big|  X \Big) =
\begin{cases}
   constant & \text{if } X = 
  \bar S\,,\\
    \int_{ S} d M_{\rm hm} \P \Big(d \Big| M_{\rm hm} \Big) ~ \times \P\Big(M_{\rm hm} \Big|  S\Big)  & \text{if }   X =  S\,.
\end{cases}
   \end{multline}  
   We choose a log-uniform prior distribution $ \P\Big(M_{\rm hm} \Big|  S\Big)$ on $M_{\rm hm}$ within $S$, which corresponds to a prior that is non-informative about the order of magnitude of the half-mode mass. 
   We obtain the $constant$ and $\P \Big(d \Big| M_{\rm hm} \Big)$ by dividing the posterior of the original analysis by its prior. We further enforce that all probabilities add up to 1 in the posterior in order to obtain the correct normalisation.
   
   This framework allows us to include the idealised CDM case while maintaining the log-uniform prior regarding the sensitive region. It comes, however, at the small cost that we can only report an upper limit in the case that it happens to fall within the sensitive region.
    Our first summary is the $95$ percentile of the posterior in this hyper model scenario, $M_{\rm hm}^{C\rm L}$, whose defining equation is:
    \begin{multline} 0.95  = \P(\bar S|d) \\ + \int_{10^{5} ~{\rm M_{\odot}h^{-1}}} ^{M_{\rm hm}^{\rm CL} } ~dM_{\rm hm}~ \P \Big(d \Big| M_{\rm hm} \Big) \times \P\Big(M_{\rm hm} \Big|  S\Big) \frac{\P(S)}{\P(d)}\,.
    \end{multline}

  In the equation above, $\P(S)$ is the prior probability of the sensitive case. The original parameter range contains all half-mode masses between the one corresponding to the coldest WIMP model and the constraints from the cosmic microwave background.
  For a log-uniform prior on half-mode masses, this corresponds to  $\P(S)=1-\P(\bar S)=0.45$.
  We use this prior when reporting upper limits in this section for simplicity, but in general, one could choose different prior values. In Fig. \ref{fig:prior_dep}, we show how the 95 percentiles on the half-mode scale change as a function of prior mass attributed to the sensitive region. We find that the order of magnitude of these 95 percentiles is stable for values of $\P(S)$ between $0.5$ and $1.0$.

    Following this approach we find a joint upper limit of $ \lambda_{\rm hm}^{\rm CL} =  0.089 {\rm~ Mpc~ h^{-1}}$.
   This rules out that haloes with a mass of $M_{\rm hm}^{\rm CL} = 3\times 10^{7 }$~${\rm M_{\odot}~ h^{-1}} $ are significantly suppressed with respect to the CDM scenario at the $2\sigma$ level. 
    Under the assumptions discussed in Section \ref{sec:dm_model_data}, we can express our constraints in terms of a lower limit on the thermal relic particle mass, i.e. $m_{\rm th}^{\rm CL}  =  6.048 ${\rm~keV} at the 95 per cent confidence level.
    We mark these limits with dashed vertical lines in Figure  \ref{fig:thermal_part_mass}.
    These constraints are in agreement with those derived by previous studies, as summarized in Table \ref{tab:comp}. We find that we require a higher sensitivity towards lower halo masses in order to rule out or confirm CDM models. 
    Notice that our model assumptions, for example, on the IGM priors in the Lyman-$\alpha$ analysis (see section \ref{sec:systematics_lyman}), are rather conservative. While we obtain mildly weaker limits with respect to past literature, our limits are expected to be more robust.

 \subsubsection{Bayes factors} 
 \label{sec:statsum_BF}
\begin{figure}
\includegraphics[width=1.0\hsize]{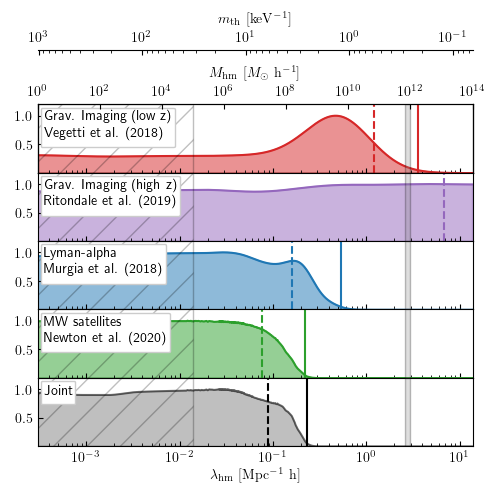}
 \caption{The posterior probability distributions for the analysis of the gravitational lensing analysis of extended arcs for the SLACS sample (red) and the BELLS sample (purple), the Lyman-$\alpha$ forest data (blue) and the luminous satellites of the MW (green).  All posteriors are scaled so that their maximum value is 1. The grey hatched area highlights the region in which all of the probes considered here become insensitive to the difference between the different models. The mass of the MW within the $68$ per cent confidence interval, as inferred by \citet{Callingham_2019}, is shown  with a grey line at $M_{\rm hm} \approx 10^{12}{\rm M}_{\odot} \approx M_{200}^{\rm MW}$.
 The vertical (dashed) lines indicate the upper limits determined from the Bayes factor ( the $95$ percentile) criterium.
 }
\label{fig:thermal_part_mass}
\end{figure}

\begin{figure}
	\includegraphics[width=1.0\hsize]{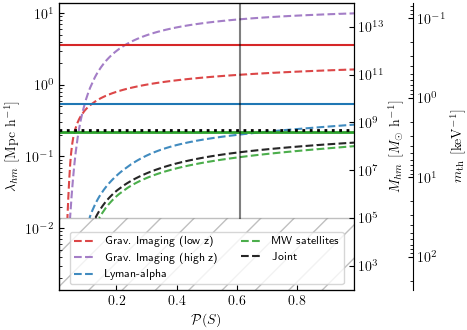}
    \caption{ The behaviour of 95 per cent upper c.l. (dashed curves) as a function of the prior mass attributed to the sensitive region. The hatched area highlights the region in which none of the probes considered here is sensitive anymore. The vertical line shows the prior $\P(S)$ corresponding to the original box in which the analyses were performed (see Table \ref{tab:posterior_summaries}). Notice that the order of magnitude of 95 percentiles is stable over a large range of values for these probes. For reference we show the value of the  upper limit according to the Bayes factor criterium for: the joint posterior (dotted black), the Lyman-$\alpha$ forest posterior (solid blue), the Milky Way satellites posterior (solid green), and the SLACS sample of lens systems (solid red).}
    \label{fig:prior_dep}
\end{figure}

 In order to be less dependent on the chosen parameter range and prior assumptions, the second summary statistic considers the ratio of likelihood with a model $\lambda_{\rm hm}$ and the model that maximises the likelihood $\lambda_{\rm hm}^{\rm ML}$ (corresponding to the Bayes factor between these two models, when each parameter value is considered to be different model). 
    The value $\lambda_{\rm hm}^{\rm BF}$, above which the ratio of all models fullfill $\frac{\P(d|\lambda_{\rm hm}>\lambda_{\rm hm}^{\rm BF}) }{\P(d|\lambda_{\rm hm}^{\rm ML})}  \leq \frac{1}{20}$  gives then an upper limit in the sense that all these models are strongly disfavoured (i.e. ruled out at 95\% confidence limit) in comparison to the maximum likelihood case. We mark these upper limits with solid vertical lines in Figure  \ref{fig:thermal_part_mass}.
    
    We find for the joint posterior an upper limits of $\lambda_{\rm hm}^{\rm BF} = 0.233 {\rm ~Mpc ~ h^{-1}}$, corresponding  to $M_{\rm hm}^{\rm BF} = 4.8 \times 10^{8} {{\rm~ M_{\odot} ~ h^{-1}} }$ and $m_{\rm hm} = 2.552 {\rm~ keV}$. This upper limit is mostly determined by the analysis of MW satellites analysis, with $\lambda_{\rm hm}^{\rm BF} =  0.221 {\rm~ Mpc ~ h^{-1}}$. The Lyman alpha forest, with $\lambda_{\rm hm}^{\rm BF} = 0.540{\rm ~Mpc ~ h^{-1}}$, turns out to be the second strongest constraint.
    {
    We find that for the lensing probes only the SLACS sample exclude values according to this summary criterium, with  $\lambda_{\rm hm}= 3.607 {\rm~ Mpc ~ h^{-1}}$. In the case of the BELLS-GALLERY, the posteriors actually prefer the warmer dark matter models. This is reflected in the ratio between the maximum likelihood value and the likelihood of the cold limit, which is 1/1.14 at $\lambda_{\rm hm}^{\rm ML}  = 4.538{\rm ~ Mpc ~ h^{-1}}$ for \citetalias{ritondale2019}, respectively.}
    We summarize the different results in Table \ref{tab:posterior_summaries}, which furthermore gives additional information about the individual probes.

\section{Systematic errors}
\label{sec:systematics}

In this section, we discuss the different sources of systematic errors that may affect each of the astrophysical probes considered here.

\subsection{Strong gravitational lensing}
\label{sec:systematics_lensing}

The main sources of systematic errors that are common to strong gravitational lensing techniques are related to the assumptions on the mass density profile of the main lenses and their subhaloes, and the normalisation of the halo mass function.

\subsubsection{Departures from power-law mass models}
In the context of strong gravitational lensing by galaxies, the standard procedure is to parameterise the mass distribution of the lens with an elliptical power-law profile and a contribution of an external shear component. However, both numerical simulations \citep{Xu_2015,Hsueh_2018_ill} and observations \citep[][]{Gilman_2017,Hsueh2016,hsueh2017,xu2013colddarkmatter}  demonstrated that for the analysis of lensed  quasars, in some cases, important departures from this simplified model exist and have a non-negligible effect on the inference of low-mass haloes. For example, \citet{Hsueh_2018_ill} showed that the presence of an additional disc component could increase the probability of finding significant flux-ratio anomalies by 10 to 20 per cent, while baryonic structures in early-type galaxies lead to an increase of the order of 8 per cent. Similar effects are expected for departures in the mass distribution from a power-law in early-type galaxies in the analysis of extended sources \citepalias[][]{ritondale2019}. However, for a potentially different conclusion see \citet{enzi2020}. Both \citetalias{vegetti2018} and \citetalias{ritondale2019} explicitly avoid this problem by including pixellated corrections to the lensing potential. These corrections are used to detect the low-mass haloes themselves and to distinguish them from other forms of complexity, i.e. they can also be used to account for large-scale deviations from the assumed elliptical power-law mass model \citep{vegetti2014}.

\subsubsection{Normalization of the mass functions} 

Numerical simulations have shown that the normalisation of the subhalo mass function depends on the virial mass and redshift of the lens galaxy \citep[see e.g.][]{Xu_2015,Gao_2010}. 
Including this evolution becomes critical when analysing heterogeneous samples of lenses, such as those studied by \citet{gilman2019_newest} and \citet{hsueh2019}. 
However, this can be a challenging task. First, strong gravitational lensing only provides a measure of the projected mass within the Einstein radius and deriving a virial mass requires extrapolations of the lens model and the use of empirically calibrated relations, such as those between stellar mass and virial mass \citep[e.g.][]{Auger_2010b,Auger_2010,Sonnenfeld_2018}. Second, the evolution of the subhalo mass function with host redshift and mass depends on the accretion history and the survial rate of the accreted subhaloes. 
For example, as briefly discussed in Section \ref{sec:marginal},  tidal interactions can lead to the disruption of subhaloes and affect both the normalisation of their mass function of and their spatial distribution  \citep[see e.g.][]{Hayashi_2003,Gao_2004,Green_2021}. Baryonic physics can enhance this effect, particularly in the innermost $\sim 50$ kpc of the host halo close to the central galaxy \citep[see e.g.][]{sawala_shaken_2017}, in a way that depends on the detailed implementation of the galaxy formation model and the physics of dark matter.

Due to these complications, we have not explicitly included a dependency of the subhalo mass function with the host virial mass (though we include a dependence on the host mass within twice the Einstein radius) and redshift. While we plan to include this effect in a follow-up publication, we do not expect this assumption to affect our current results significantly for the following reasons:
\begin{enumerate}
\item due to their selection functions, both the SLACS and the BELLS-GALLERY samples span a narrow range in redshift. 
\item By marginalizing over the normalisation constant of the subhalo mass function (or equivalently $f_{\rm sub}^{\rm CDM}$) before multiplying the posteriors, we obtain constraints on $\lambda_{\rm hm}$ that are not affected by the difference in the mean redshift of the two samples. 
\item Our allowed range of normalization constants is consistent with the one by \citet{gilman2019_newest} which have included the dependence on the host virial mass and redshift more explicitly.
\end{enumerate}

We have also assumed that the line-of-sight halo mass function has a fixed normalisation equal to the mean normalisation value from CDM numerical simulations.
\citet{Treu2009_non_overdense} have shown that the line of sight of the SLACS lenses have densities that are comparable to those of non-lensing early-type galaxies with a similar redshift and mass. However, massive galaxies may preferentially reside in lines of sights that are systematically over-dense \citep[see][for differing results]{Fassnacht_2010,Collett_2016}, which could bias our results towards colder dark matter models. Moreover, the typical line of sight for different WDM models may not be the same as for the CDM case.
This effect is potentially problematic for the analyses by \citetalias{vegetti2018} and \citetalias{ritondale2019}, as their samples of lenses are homogeneous and consist of massive early-type galaxies. 
As for the subhalo mass function normalisation, high-resolution and large volume numerical simulations in CDM and several WDM models are the key to shed light on these issues.

\subsubsection{Low-mass halo profiles}

\begin{figure}
\includegraphics[width=1.0\hsize]{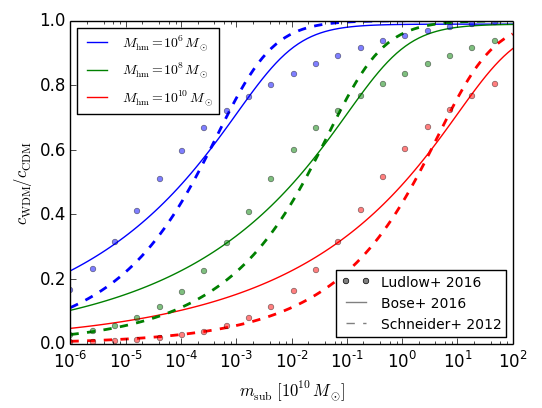}
 \caption{  The concentration of WDM haloes relative to CDM predictions, for three different half mode masses. We show changes in the concentration as a function of the halo mass determined with the relations of \citet[][]{schneider2012}, \citet{bose_copernicus_2016} and \citet{ludlow2016}  at z=0.5. We note that these relations have originally been fitted to halo masses $\gtrsim10^{9}M_\odot$ and their application to the lower-mass haloes requires some extrapolation. 
 }
\label{fig:CEvo}
\end{figure}

We assume that both haloes and subhaloes are well described by a spherical NFW profile that follows the concentration-mass-redshift relation of \citet{duffy2008}. However, this assumption has some drawbacks:
\begin{enumerate}
    \item due to non-linear processes, such as tidal stripping by the host halo, one expects the profiles of subhaloes and, in particular, their concentration to change as a function of distance from the host centre \citep{Moline_2017}.
    A tidally stripped subhalo will be more concentrated than a field halo of the same mass, making it easier to be detected \citep[see e.g.][]{Quinn2020}. Hence, our assumption that we can neglect these processes reduces the number of detectable objects in our analyses \citep[below 10 per cent][]{despali2018} and renders our results conservative. 
    Moreover, \citet{despali2018} have explicitly quantified the effect of these assumptions and found it to be relatively small, underestimating the subhalo mass by at most 20 per cent.
    Indeed, the overall differences in the reconstructed subhalo mass function is a slight shift towards smaller masses, that is smaller than the intrinsic uncertainty of reconstructed masses among different subhalo finding algorithms \citep{Onions_2012}, and no significant bias in the reconstructed half-mode scale is expected.
    The above discussion shows that our assumptions are conservative and we note that the importance of tidal interactions is further mitigated by the fact that the contribution to the lensing signal from line-of-sight haloes is at least equal (but often higher) than the subhalo contribution \citep[][]{despali2018}.
    
    \item As structure formation is delayed in WDM models, it is expected that the concentration-mass-redshift relation is different than for CDM \citep[see Figure \ref{fig:CEvo} and][]{schneider2012,bose_copernicus_2016,ludlow2016}.
    On the other hand, \citet{despali2018} have shown that the difference in the lensing effect due to a change in concentration is at most of the order of 10 per cent.
    We plan to investigate this issue more thoroughly in a follow-up publication.
    \\
    \item The concentration-mass-redshift relation is typically derived from simulations with subhalo masses greater than $ 10^9 M_{\odot}$. Applying this relation to the mass range relevant for this work
    requires therefore an extrapolation of several orders of magnitudes in mass, and further highlights the need for higher-resolution (hydrodynamical) simulations that describe the evolution of low-mass haloes in CDM and WDM.
     
\end{enumerate}

\subsection{\texorpdfstring{Lyman-$\alpha$}{Lyman-alpha} forest}
\label{sec:systematics_lyman}

Here, we address the systematics affecting the analysis of the Lyman-$\alpha$ forest data by summarizing the discussion presented by
\cite{viel2013}.

One of the potential systematics arises from the box size and particle number of the numerical simulations used for the model comparison, for which different setups usually show deviations at the 5 to 15 per cent c.l. \citetalias{murgia2018} corrected for this effect in their analysis by comparing their simulations with those from standard cosmological simulations.

On small scales, the quasar spectra are influenced by the instrumental resolution, which in the case of the MIKE and HIRES data sets are at most on the level of 20 and 5 per cent, respectively. This uncertainty is independent of the redshift. The uncertainties arising from the signal-to-noise ratio of the spectra on the smallest scales vary from around 2 per cent at $z\leq 5$ to 7 per cent for the highest redshift bin. UV fluctuations in the spectra have been implemented using a rather extreme model that only takes into account the ionizing effect of the quasars. The systematic effect on flux power spectra is expected to be $\leq 10$~per cent for the scales considered here and is scale-dependent \citep{croft2004,mcdonald2006}. An additional systematic associated with the quasar spectra is the contamination with metal lines in the Lyman-$\alpha$ forest. However, this is expected to add less than 1~per cent to the uncertainty of flux power spectra on all of the scales considered in the analysis.

A well-known issue affecting Lyman-$\alpha$ forest analyses is the degeneracy between the small-scale impact of different WDM models, and the heating effects due to different thermal or reionization IGM histories (e.g., \citet{Garzilli_2017,garzilli2019warm}. Unlike the IGM temperature, the WDM mass is a redshift-independent parameter. Thus, by simultaneously fitting power spectra at different redshift bins, one can partly break the degeneracy \citepalias{murgia2018}. Furthermore, the limits presented in this work are obtained considering the IGM temperature $T_{\rm igm}(z)$ as a freely floating parameter, redshift bin by redshift bin. In other words, we did not make any assumption on its redshift-evolution, besides imposing $T_{\rm igm}(z) > 0$, and $\Delta_{\rm igm}(z) < 5000$ between adjacent redshift bins.

\subsection{Milky Way luminous satellites}

One of the major nuisance parameters affecting the constraints on dark matter obtained from the analysis of the luminous MW satellites is the mass of the MW, $M_{200}^{\rm MW}$. 
For this mass, \citetalias{newton2020constraints} 
assume a value within the current observational constraints, such as those by \citet{Callingham_2019} or \citet{wang2019}. Changes to the assumed MW halo mass alter the number of subhaloes of a given mass that host a visible galaxy \citep[see e.g.][]{sawala_shaken_2017}. 
For example, doubling the halo mass approximately doubles the number of subhaloes \citep{wang_satellite_2012,cautun_subhalo_2014}. 

The analysis by \citetalias{newton2020constraints} models galaxy formation using the approach of  \citet{lacey_unified_2016}. This model has been
calibrated extensively by comparison with the luminosity function and properties of galaxies in redshift surveys, and it agrees with the predicted CDM satellite luminosity function of the Milky Way and Andromeda \citep{bose_imprint_2018} in its standard implementation. The main process that determines the total number of Galactic satellites is reionisation, which curtails star formation in the faintest galaxies and thus sets the faint end of the dwarf galaxy luminosity function.  The predictions of \citetalias{newton2020constraints} assume a reionisation redshift of $z_{\rm reion}=7$ that is in agreement with the latest CMB measurements \citep{planck2020}, although it is at the lower end of the allowed range.  It further agrees with observations of high redshift quasars that set a lower limit for the end of the epoch of reionization being before a redshift of $z=6$ \citep[see e.g.][]{Cen_2002}. 
A later epoch of reionisation leads to more ultra-faint dwarfs.
The choice of $z_{\rm reion}=7$ is conservative, since an overprediction of the satellite luminosity function leads to stricter constraints on the half mode mass. An earlier epoch of reionisation, i.e. choosing a larger value than $z_{\rm reion}=7$, would, therefore, provide more stronger constraints on the WDM particle mass \citepalias[as has been shown by][]{newton2020constraints}.

An important systematic associated with this astrophysical probe is the choice of the observed satellite population. Half of the non-classical satellites in the sample are drawn from the SDSS and have been spectroscopically confirmed as DM-dominated dwarf galaxies.  \citetalias{newton2020constraints} draw the other half from the DES, only 25 per cent of which are spectroscopically-confirmed. If later work reclassifies some of the DES objects to be globular clusters, then the inferred total satellite count will decrease for faint objects. However, this effect is likely to be small due to the good agreement in the inferred MW satellite luminosity function when using only SDSS or only DES observations.

In their analysis,  \citetalias{newton2020constraints}  assume that the MW and its satellite system are typical examples of most DM haloes with similar masses. If this is not the case, for example, due to environmental effects, one expects that this would affect the analysis. M31, for example, could introduce anisotropies into the MW subhalo distribution. 
In general,  if the radial distribution of subhaloes in simulations is different from the distribution within the MW, it can lead to systematic uncertainties. Anisotropies would give rise to a correlation between satellites. 
\citet{newton_total_2018} briefly study the effects of anisotropy in the subhalo distribution and choose 300 kpc as their fiducial radius (smaller than the distance between MW and M31) to minimize the effects from interactions with M31.

\section{Future prospects}
\label{sec:future_prosepect}

In this section, we discuss how the current constraints from the three different probes are likely to improve in the near future, and which steps will be necessary to obtain a more precise measurement on dark matter.

\subsection{Gravitational lensing}

The level of constraints currently obtainable with strong gravitational lensing is mainly determined by the low number of known systems, in particular at high-redshift (i.e. those for which the line of sight contribution is maximal). Moreover, in the particular case of extended sources, the lack of high-angular-resolution data strongly limits the possibility of detecting haloes with masses below $10 ^8$~M$_\odot$. This hinders the exploration of the region of the parameter space, where the difference between different dark matter models is the largest.

Ongoing and upcoming surveys are expected to lead to the discovery of a large number of new gravitational lens systems. {\it Euclid}, for example, is expected to deliver as many as $\mathcal{O}(10^{5})$ new lensed galaxies \citep{Collett_2015}, while $\mathcal{O}(10^{3})$ lensed quasars are expected to be found in future surveys of the Vera C. Rubin Observatory, formerly known as the Large Synoptic Survey Telescope \citep[LSST,][]{Oguri_2010}. However, these new samples on their own will not be sufficient to significantly and robustly improve upon the present constraints. In particular, the gravitational imaging approach will require high-resolution follow-up observations to probe halo masses below the current limits. As the expected angular resolution of {\it Euclid} is about two times worse than currently available with the {\it HST} and about four times worse than what is already provided by current adaptive optics systems, these observations will only allow us to probe the halo mass function in a regime where predictions from different thermal relic dark matter models are essentially the same. 

These follow-up observations can come from extremely large telescopes, such as the  Thirty Meter Telescope (TMT), the Giant Magellan Telescope (GMT), and the European Extremely Large Telescope (E-ELT), as well as VLBI observations at cm to mm-wavelengths, which will provide an angular resolution of the order of  $\sim0.2$ to 5 milli-arcseconds. 
This will open up the possibility of detecting haloes with masses as low as $10^6$~M$_\odot$ \citep{mckean2015strong,Spingola_2018}. 
Furthermore, the {\it James Webb Space Telescope} (JWST) will not only provide an angular resolution of $\sim0.02$ to 0.1 arcsec, but will also allow us to maximise the contribution from the line of sight haloes by targeting high-redshift systems, and therefore, can potentially deliver tighter constraints on the mass function in the mass ranges currently probed. 

We notice that flux-ratios of gravitationally lensed quasars also pose a very promising probe of dark matter. In order to take full advantage of their observations, deep follow-up imaging will be needed to quantify the frequency of galactic discs and other forms of complexity in the lens mass distribution, while long term monitoring will provide a robust measurement of the relative fluxes and possible variability in the lensed images \citep[][]{harvey2019,koopmans2003}. It should also be considered that higher angular-resolution observations of such systems will allow us to resolve the extended source structure, and, therefore, permit an analysis using the gravitational imaging approach.

With increasing resolution and sample sizes, fully understanding all sources of systematic errors will become increasingly important. To this end, high-resolution, realistic hydrodynamical simulations in different dark matter models will be required (e.g. \citealt{mukherjee2018,enzi2020}).

\subsection{\texorpdfstring{Lyman-$\alpha$}{Lyman-alpha} forest}

In the near future, more accurate measurements of the IGM thermal history will provide stronger priors for the data analyses, allowing us to better constrain the small-scale cut-off in the linear power spectrum \citep[see e.g.][]{boera2018}.

Furthermore, the inclusion of the set of intermediate resolution and signal-to-noise quasar spectra observed by the XQ-100 survey \citep{lopez2016}, and of the new, high-resolution ones observed by the ESPRESSO spectrograph \citep{pepe2019},  will improve the constraints presented in \citetalias{murgia2018} and in this work,
due both to improved large number statistics and to the complementary redshift and scale coverage, which will break some of the degeneracies among different parameters. 

Another possible refinement might be achieved by including additional hydrodynamical simulations for which both astrophysical parameters, e.g., the IGM temperature, and WDM mass, are varied simultaneously. Constraints from the Lyman-$\alpha$ forest on small scales are indeed limited by the thermal cut-off in the flux power spectrum introduced by pressure and thermal motions of baryons in the ionised IGM. This makes the determination of accurate and independent constraints on the IGM thermal history essential in order to push current limits to even larger thermal relic masses. The 21 cm signal from neutral hydrogen gas before reionisation could provide such an independent measurement 
\citep[see e.g.][]{viel2013}.

Concurrently with ongoing and future experimental efforts, further theoretical work is thus needed to interpret observations, accurately disentangle the impact of the various parameters, and combine outcomes from different observational methods.

\subsection{Milky Way luminous satellites}

There are two aspects of Local Group studies that are expected to improve in the future. The first relates to the theoretical predictions as simulations improve, and the second comes from the improving observational data as larger and deeper surveys are carried out, and new detection methods are developed.

Subhaloes in simulations can only be resolved above a certain particle number, which results in missing low-mass subhaloes. This issue can be approximately corrected for; however, future high-resolution simulations may lower the mass scale below which one needs to make these corrections.
Also, next-generation simulations will assist attempts to understand better the relevant (baryonic) processes of satellite formation, potentially opening up the possibility to not just present an upper limit on $M_{\rm hm}$ from the abundance of MW luminous satellites.

The method used by \citetalias{newton2020constraints}  \citep[based on][]{newton_total_2018}
assumes that the observed satellites, which are found in surveys with various detectability limits, are a representative sample of the global population. However, there could be a population of faint and spatially extended dwarfs that are inaccessible to current surveys \citep[see e.g.][]{Torrealba_2016a,Torrealba_2016b}. 
The WDM constraints inferred from the satellite distribution could be improved further by deep observations of other nearby galaxies besides the MW, such as M31, Centaurus A or the Virgo Cluster. Such external observations help to reduce uncertainties in the current analysis arising from the MW halo mass and from the halo-to-halo scatter of the satellite luminosity function.

Finally, stronger limits on the halo mass of the MW and especially the Large Magellanic Cloud (LMC) could help to provide a better model of the satellite number counts, as the LMC is known to have brought its own satellites \citep{kallivayalil2018,Patel2020} that need to be properly accounted for \citep{Jethwa2016}.

\section{Conclusions}
\label{sec:conclusion}

We have derived new constraints on thermal relic dark matter models from the joint statistical analysis of a set of different astrophysical probes. In particular, we extended two previous studies of strong gravitational lens systems and combined them with constraints from the Lyman-$\alpha$ forest and the luminous MW satellites. Our results have interesting implications for the current status of dark matter studies, their limitations, as well as the most promising ways to improve upon them in the near future.  We summarize them as follows: 

\begin{enumerate}

\item We determined limits by considering the 95 percentiles of the parameters describing WDM models. From our joint posterior we find a upper limit on the half-mode scale of $\lambda_{\rm hm}^{\rm CL} =0.089{\rm~Mpc ~ h^{-1}}$, corresponding to $M_{\rm hm }^{\rm CL}=3 \times 10^{7}{\rm~ M_{\odot} ~ h^{-1}}$ and a lower limit of $m_{\rm th}^{\rm CL}  = 6.048${\rm~keV} under the assumption of Planck cosmology and a thermal relic dark matter model.
 These limits rule out the $7.1 {\rm ~ keV}$ sterile neutrino dark matter model for a lepton asymmetry $L_6>10$. If such sterile neutrino models aim to explain the observed $3.55 {\rm ~ keV}$ they are required to show a half-mode mass in the range of $\log_{10}(M_{\rm hm} \cdot {\rm M_{\odot}^{-1}~ h}) \in [9,11]$.
According to this summary, we furthermore rule out the ETHOS-4 model of self-interacting DM, which shows a cutoff corresponding to a thermal relic with a mass $m_{\rm th} = 3.66 {\rm ~keV}$ \citep{Vogelsberger_2016}. 
Amongst the considered probes, the MW satellites and  the Lyman-$\alpha$ forest provide the strongest constraints on the half-mode scale, i.e. $\lambda_{\rm hm}^{\rm CL}<0.076 {\rm~ Mpc~ h^{-1}}$, and $\lambda_{\rm hm}^{\rm CL}< 0.160 {\rm~ Mpc~h^{-1}}$, respectively. These values are followed by the  strong gravitational lensing constraints of the SLACS sample $\lambda_{\rm hm}^{\rm CL}< 1.214 {\rm ~Mpc~h^{-1}}$, and the weakest constraints coming from the high redshift BELLS-GALLERY and $\lambda_{\rm hm}^{\rm CL}< 6.842 {\rm ~Mpc~h^{-1}}$.
The latter even shows a preference for warmer dark matter models, in contrast to the other probes. However, larger samples and higher-sensitivity lensing data are required to confirm such a trend.

\item We further considered the ratios of the joint likelihood, we find that with respect to the maximum likelihood model, we rule out models above $ \lambda_{\rm hm}^{\rm BF}=0.233 {\rm~ Mpc~h^{-1}}$ (corresponding to values above  $M_{\rm hm}^{\rm BF}= 4.8 \times {10^{8} {\rm ~ M_{\odot}~ h^{-1}}}$ and below $m_{\rm th}^{\rm BF}=2.552 {\rm ~ keV}$). 
 Again, we find that the sterile neutrino dark matter models are ruled out. However, due to weaker constraints, the self-interacting DM  model of ETHOS-4 is still allowed.
In the case of Bayes factors, the limits are again mostly determined by the analysis of the Milky Way satellites (with  $ \lambda_{\rm hm}^{\rm BF}=0.221 {\rm~Mpc~h^{-1}}$). The Lyman-$\alpha$ analysis follows with  $ \lambda_{\rm hm}^{\rm BF}=0.540 {\rm ~Mpc~h^{-1}}$. In the case of lensing probes, only the SLACS sample provides an upper limit under this criterium. We find an upper limit of $ \lambda_{\rm hm}^{\rm BF}=3.607 {\rm~Mpc~h^{-1}}$ in this case.

\item We highlight that the choice of a summary statistics is crucial for deciding which dark matter models are ruled out. In general, we find that the 95 percentiles provide stronger constraints, while the Bayes factor summary statistics provide more conservative limits (that are also more independent from prior assumptions).

\item None of the considered analyses are sensitive to half-mode masses below $M_{\rm hm} = 10^5 {\rm ~M_{\odot}~ h^{-1} } $, where the likelihood and posterior distributions flatten out. In the near future, we expect strong lensing observations with extended sources to increase their sensitivity towards these colder models thanks to the improvement in the angular resolution that will be provided by VLBI and the ELTs. High spectral resolution observations of quasars will provide Lyman-$\alpha$ forest constraints on smaller scales of the matter power spectrum and, therefore, smaller values of $\lambda_{\rm hm}$ \citep{irsic2017}. For both probes, a larger sample of objects is expected to lead to more precise constraints. An analysis of the luminous MW satellites, on the other hand, is by definition limited to satellites that are massive enough to host a galaxy. This restriction puts a limit on the lowest subhalo mass that can be detected, and the relative constraints will only improve with better control of systematic errors.

\item All probes are affected by their model assumptions (Section \ref{sec:methods_data})   and different sources of systematic errors (Section \ref{sec:systematics}) that will need to be addressed to improve on the current level of accuracy. It is a well-known fact that current observations of the Lyman-$\alpha$ forest can be compatible with both CDM and WDM, depending on the assumptions made on the thermal history of the IGM.
The interpretation of the MW satellite luminosity function is strongly affected by poorly constrained feedback and star formation processes, as well as the mass of the MW (\citealt{lovell2012,lovell_properties_2014} and references therein). Inference on the halo and subhalo mass function from strong lensing observations can be significantly biased by assumptions made on the lens mass distribution  \citep[for both lensed galaxies and quasars,][]{vegetti2014} and the size of the background sources (mainly for lensed quasars; Timerman et al., in prep.).

\end{enumerate}

In this paper, we have focused on three different astrophysical observations to place constraints on thermal relic dark matter, strong gravitational imaging, the Lyman-$\alpha$ forest, and the luminosity function of the MW satellites.
One of the major opportunities of a joint analysis of different astrophysical observations is the possibility to correct biases present in the individual posteriors. We note, however, that this may lead to joint constraints which are weaker than those of each individual probe.

In the future our study could be extended by considering the number of non-luminous MW subhaloes detectable with stellar streams \citep[][]{Banik_2018_naturedm,banik2019novel,Carlberg_2012,Carlberg_2013,erkal2015,Yoon_2011}, the analysis of flux ratios of multiply lensed quasars \citep[see e.g.][]{metcalf2001,hsueh2019,gilman2019,gilman2019_newest}, the cosmic microwave background \citep{planck2016}, the luminosity function of satellites in galaxies other than the MW \citep{Nierenberg_2011,Nierenberg_2012,Corasaniti_2017}, and  the constraints from the observed phase-space density of dwarf spheroidal galaxies \citep[see e.g.][]{alvey2020new}.
Further probes of WDM can be found in the cosmic reionization and gravitational waves \citep{Tan_2016} and the number density of direct-collapse black holes \citep{Dayal_2017}.
Finally, one could extend the study presented in this work to other alternative dark matter models that affect the power spectrum and, therefore, the mass function of haloes. Examples of such models are fuzzy, 
and potentially self-interacting dark matter.

\section*{Acknowledgements}

SV thanks the Max Planck Society
for support through a Max Planck Lise Meitner Group, and
acknowledges funding from the European Research Council (ERC)
under the European Union’s Horizon 2020 research and innovation
programme (LEDA: grant agreement No 758853). CSF acknowledges support from the European Research Council through the ERC Advanced Investigator grant, DMIDAS [GA 786910]. JPM acknowledges support from the Netherlands Organization for Scientific Research (NWO) (Project No. 629.001.023) and the Chinese Academy of Sciences (CAS) (Project No. 114A11KYSB20170054). CDF acknowledges support for this work from the National Science Foundation under Grant No. AST-1715611. MRL acknowledges support by a Grant of Excellence from the Icelandic Research Fund (grant number 173929). MC acknowledges support by the EU Horizon 2020 research and innovation programme under a Marie Sk{\l}odowska-Curie grant agreement 794474 (DancingGalaxies). RM acknowledges the computational resources provided by the Ulysses SISSA/ICTP supercomputer. ON was supported by the Science and Technology Facilities Council (STFC) through grant ST/N50404X/1and acknowledges support from the Institute for Computational Cosmology (ICC) PhD Scholarships Fund and thanks the benefactors who fund it.  ON also acknowledges financial support from the Project IDEX-LYON at the University of Lyon under the Investments for the Future Program (ANR-16-IDEX-0005) and supplementary financial support from La R\'{e}gion Auvergne-Rh\^{o}ne-Alpes. This work was also supported by STFC Consolidated Grants for Astronomy at Durham ST/P000541/1 and ST/T000244/1. This work used the DiRAC Data Centric system at Durham University, operated by the Institute for Computational Cosmology on behalf of the STFC DiRAC HPC Facility (www.dirac.ac.uk). This equipment was funded by BIS National E-infrastructure capital grants ST/K00042X/1, ST/P002293/1, ST/R002371/1 and ST/S002502/1, Durham University and STFC operations grant ST/R000832/1. DiRAC is part of the National e-Infrastructure.

\section*{Data Availability}
The derived data generated in this research will
be shared on reasonable request to the corresponding author.




\bibliographystyle{mnras}
\bibliography{paper} 



\appendix


\bsp	
\label{lastpage}
\end{document}